\documentclass[%
 reprint, amsfonts,amssymb,amsmath,longbibliography,
superscriptaddress,
 amsmath,amssymb,
 aps,
prb,
]{revtex4-2}

\usepackage{graphicx}
\usepackage{dcolumn}
\usepackage{bm}
\usepackage[colorlinks=true,citecolor=blue,linkcolor=blue,urlcolor=blue]{hyperref}
\usepackage{color}
\usepackage{upgreek}

\newcommand{\bfq}{\mathbf{q}}

\begin{document}


\title{Investigation of spin excitations and charge order in bulk crystals of the infinite-layer nickelate LaNiO$_2$}

\author{S.~Hayashida}
\email[]{s.hayashida@fkf.mpg.de}
\affiliation{Max-Planck-Institute for Solid State Research, Heisenbergstra$\beta$e 1, 70569 Stuttgart, Germany}
\author{V.~Sundaramurthy}
\affiliation{Max-Planck-Institute for Solid State Research, Heisenbergstra$\beta$e 1, 70569 Stuttgart, Germany}
\author{P.~Puphal}
\affiliation{Max-Planck-Institute for Solid State Research, Heisenbergstra$\beta$e 1, 70569 Stuttgart, Germany}
\author{M.~Garcia-Fernandez}
\affiliation{Diamond Light Source, Harwell Campus, Didcot, Oxfordshire OX11 0DE, United Kingdom}
\author{Ke-Jin~Zhou}
\affiliation{Diamond Light Source, Harwell Campus, Didcot, Oxfordshire OX11 0DE, United Kingdom}
\author{B.~Fenk}
\affiliation{Max-Planck-Institute for Solid State Research, Heisenbergstra$\beta$e 1, 70569 Stuttgart, Germany}
\author{M.~Isobe}
\affiliation{Max-Planck-Institute for Solid State Research, Heisenbergstra$\beta$e 1, 70569 Stuttgart, Germany}
\author{M.~Minola}
\affiliation{Max-Planck-Institute for Solid State Research, Heisenbergstra$\beta$e 1, 70569 Stuttgart, Germany}
\author{Y.-M.~Wu}
\affiliation{Max-Planck-Institute for Solid State Research, Heisenbergstra$\beta$e 1, 70569 Stuttgart, Germany}
\author{Y.~E.~Suyolcu}
\affiliation{Max-Planck-Institute for Solid State Research, Heisenbergstra$\beta$e 1, 70569 Stuttgart, Germany}
\author{P.~A.~van~Aken}
\affiliation{Max-Planck-Institute for Solid State Research, Heisenbergstra$\beta$e 1, 70569 Stuttgart, Germany}
\author{B.~Keimer}
\email[]{b.keimer@fkf.mpg.de}
\affiliation{Max-Planck-Institute for Solid State Research, Heisenbergstra$\beta$e 1, 70569 Stuttgart, Germany}
\author{M.~Hepting}
\email[]{hepting@fkf.mpg.de}
\affiliation{Max-Planck-Institute for Solid State Research, Heisenbergstra$\beta$e 1, 70569 Stuttgart, Germany}

\date{\today}

\begin{abstract}
Recent x-ray spectroscopic studies have revealed spin excitations and charge density waves in thin films of infinite-layer (IL) nickelates. However, clarifying whether the origin of these phenomena is intrinsic to the material class or attributable to impurity phases in the films has presented a major challenge. Here we utilize topotactic methods to synthesize bulk crystals of the IL nickelate LaNiO$_2$ with crystallographically oriented surfaces. We examine these crystals using resonant inelastic x-ray scattering (RIXS) at the Ni $L_3$-edge to elucidate the spin and charge correlations in the bulk of the material. While we detect the presence of prominent spin excitations in the crystals, fingerprints of charge order are absent at the ordering vectors identified in previous thin-film studies. These results contribute to the understanding of the bulk properties of LaNiO$_2$ and establish topotactically synthesized crystals as viable complementary specimens for spectroscopic investigations. 
\end{abstract}

\maketitle



\section{Introduction}
Unconventional superconductors, including cuprates, iron pnictides, and heavy fermion systems, are at the forefront of condensed-matter research. 
While the microscopic mechanisms behind their superconductivity remain a subject of debate, spin and charge instabilities  are recurrent features in their phase diagrams~\cite{ScalapinoRevModPhys2012,KeimerNat2015,FernandesNature2022,GegenwartNatPhys2008,SiScience2010}. 
The magnetic phase is typically most pronounced in the parent materials, while charge carrier doping and/or external pressure suppress the long-range magnetic order and facilitate the emergence of superconductivity \cite{Lee2006}. This suppression of long-range magnetic order transforms well-defined spin-wave excitations (magnons) of the parent materials into heavily damped excitations (paramagnons) ~\cite{BirgeneauJPSJ2006,DaiRevModPhys2015,SmidmanRevModPhys2023}, which 
can persist over wide regions of the phase diagrams~\cite{LeTaconNatPhys2011,DeanNatMat2013,MinolaPRL2015,PengPRB2018}. In cuprates, the spin excitations arise from the two-dimensional CuO$_2$ planes, exhibiting a notable bandwidth of several hundred meV~\cite{PengNatPhys2017,ColdeaPRL2001,RobartsPRB2021,Lee2014} due to strong exchange couplings ($J$), and are a prime candidate for the superconducting pairing glue~\cite{ScalapinoRevModPhys2012}. 

In addition, the ubiquitous charge density wave (CDW) ordering phenomenon might have a profound impact on the emergence of high-temperature conductivity in cuprates. These density waves are characterized by periodic electron density modulations within the CuO$_{2}$ planes, which compete with the superconducting state, at least in the underdoped regime~\cite{GhiringhelliScience2012,ChangNatPhys2012,BlancoPRB2014,LiPNAS2020}. A more complex interplay between superconductivity and various types of short-range and dynamic charge fluctuations has been observed at higher doping levels, and is a current focus of  research on the cuprates~\cite{Frano2020,ArpaiaJPSJ2021,LeeNatPhys2021,LeeCDW2021}.

In this context, recent reports of charge order in the undoped to lightly doped regime ~\cite{TamNatMat2022,KriegerPRL2022,RossiNatPhys2022,Rossi2023} of the phase diagram of infinite-layer (IL) nickelates superconductors with composition (La,Pr,Nd)$_{1-x}$Sr$_x$NiO$_{2}$~\cite{LiNat2019,ZengPRL2020,LiPRL2020,OsadaPRM2020,GaoCPL2021,ZengSciAdv2022} were not entirely unexpected, given their numerous shared basic characteristics with cuprate superconductors~\cite{BenckiserNatMat2022}. In particular, their structural motif of square-lattice NiO$_2$ planes along with a nominal 3$d^9$ electronic configuration draws close parallels to cuprates \cite{Anisimov1999}. However, this class of materials also features significant distinctions~\cite{HeptingFront2021}, such as a lack of the cuprate-typical $3d$-$2p$ orbital hybridization between the transition metal ion and oxygen. Instead, rare-earth $5d$ states contribute to the low-energy electronic structure of IL nickelates~\cite{HeptingNatMat2020,GoodgePNAS2021,RossiPRB2020}. Moreover, long-range antiferromagnetic order is absent in the parent compounds of IL nickelates~\cite{Hayward1999,Hayward2003,Crespin2005,WangPRM2020,OrtizPRR2022,KlettFront2022,FowlieNatPhys2022,Puphal2022}, although they display dispersive magnetic excitations~\cite{LuScience2021,TamNatMat2022,KriegerPRL2022,BenckiserNatMat2022,Worm2023} that are reminiscent of paramagnons in doped cuprates~\cite{LeTaconNatPhys2011,DeanNatMat2013}. As a consequence, fundamental questions about the nature of spin excitations in nickelate superconductors have remained open. 

Yet, the CDW phenomenon in IL nickelates is also debated controversially \cite{Pelliciari2023,Tam2023,RajiSmall2023,Parzyck2024}. One set of studies reported charge order in NdNiO$_{2}$ thin films without a capping layer~\cite{TamNatMat2022,KriegerPRL2022}, while the corresponding signal at approximately $\mathbf{q}=(\pm 1/3,0)$ was absent in capped films. By contrast, spin-wave excitations materialized in films with capping layer ~\cite{KriegerPRL2022,LuScience2021}, but were elusive in uncapped films~\cite{TamNatMat2022,KriegerPRL2022}. On the other hand, a different study reported the simultaneous presence of both charge order and spin excitations in LaNiO$_{2}$ films with capping~\cite{RossiNatPhys2022}, while charge order signal was not detected in PrNiO$_{2}$ and NdNiO$_{2}$ films \cite{Rossi2023}. Along these lines, recent experiments using scanning transmission electron microscopy (STEM) and x-ray scattering have cast doubt on the intrinsic origin of the CDW~\cite{RajiSmall2023,Parzyck2024}. One proposed explanation for the observed charge ordering signal involves an impurity phase, in which excess oxygen ions fill certain vacant apical positions in the IL structure, aligning in a pattern that is commensurate with a $\mathbf{q}=(\pm 1/3,0)$ ordering vector. 

To examine if previously reported spin and charge instabilities in thin-film samples originated from external factors, including interfacial effects with capping layers, oxygen off-stoichiometries, and/or damages from the invasive topochemical synthesis, we focus here on bulk crystals of the IL nickelate LaNiO$_{2}$. 
These bulk specimens with mechanically polished surfaces offer a contrasting approach to thin films, as the material exposed after polishing was neither affected by the invasive topotactic reduction atmosphere nor by contact with a capping layer. Using resonant inelastic x-ray scattering (RIXS)  on a polished LaNiO$_{2}$ surface, we detect low-energy spectral features that are attributable to spin excitations. 
We demonstrate that the observed spin excitation spectrum can be qualitatively reproduced with linear spin-wave theory and the magnetic exchange parameters known from thin films and that the lack of apparent dispersion is a consequence of the presence of three twin domains in our crystals. 
Hence, we conclude that the underlying spin excitation phenomenology in LaNiO$_{2}$ crystals aligns with that of thin films. Conversely, a CDW signal at the ordering vectors reported in previous thin-film studies is not observed in our RIXS experiment on LaNiO$_{2}$ crystals.


\section{Materials and methods}
\label{Methods}

Single crystals of perovskite LaNiO$_3$ were grown by the high-pressure optical floating zone (OFZ) method, as described in Refs.~\cite{PuphalPRM2023,PuphalAPL2023}. The as-grown centimeter-sized crystals were oriented with x-ray Laue diffraction and cut into smaller cube-shaped crystals with a wire saw, such that each surface of a cube corresponds to a pseudocubic (100) plane. The lateral dimensions of the surfaces of the cut crystals were approximately $0.8 \times 0.8$ mm$^2$.

For the topotactic oxygen deintercalation, several cube-shaped LaNiO$_3$ crystals with a total mass of about 100~mg were placed in direct contact with roughly 250~mg of powder of the reducing agent CaH$_{2}$ in a DURAN glass tube. The glass tube was evacuated to $\sim$10$^{-7}$~mbar and sealed. The reduction process was initiated by heating to 300$^{\circ}$C in a low-temperature furnace. The temperature was maintained for a duration of approximately 13~days, which is comparable to the conditions used for previous reductions of nickelate single crystals ~\cite{PuphalSciAdv2021,PuphalPRM2023}. 
The surfaces of the obtained LaNiO$_{2}$ crystals were polished using a diamond suspension, removing the uppermost layers ($\sim$30~$\upmu$m) to expose a pristine surface free of residues and secondary phases that formed during the invasive topotactic reduction.

The reduced crystals were examined by STEM, which is a sensitive method for detecting impurity phases, stacking faults in the crystal lattice, and residual oxygen in nickelates \cite{Suyolcu2021,PuphalSciAdv2021,Wu2023,Hepting2018}. Electron-transparent  specimens were prepared on a Thermo Fisher Scios I focused ion beam (FIB) using the standard liftout method. The lateral dimensions of the specimens were 20 $\upmu$m by 1.5 $\upmu$m with thicknesses between 50 and 100 nm. The STEM studies were performed with a JEOL JEM-ARM200F scanning transmission electron microscope equipped with a cold-field emission electron source, and a probe abberation-corrected Cs corrector (DCOR, CEOS GmbH) at 200 kV. STEM imaging was performed with probe semiconvergence angles of 20 mrad, resulting in probe sizes of 0.8 {\AA}. The collection angles for high-angle annular dark-field (HAADF) and annular bright field (ABF) STEM images were 75 to 310 and 11 to 23 mrad, respectively. To improve the signal-to-noise ratio of the data while minimizing sample damage, a high-speed time series (2 $\upmu$s per pixel) was acquired and then aligned and summed. 

To investigate the microstructural properties of the LaNiO$_{2}$ crystals, such as the twin domain distribution, we utilized a Zeiss Merlin electron microscope equipped with electron backscatter diffraction (EBSD) capabilities from Oxford Instruments (Symmetry EBSD detector). The AZtec and AZtecCrystal software were used for data acquisition and analysis, respectively. For the EBSD measurements, the crystal was tilted to a high angle ($70^\circ$). The detector was positioned at a 5$^\circ$ angle relative to the untilted crystal surface. Probe currents ranged 1-5 nA, and the accelerating voltage was 20 kV for most measurements. Further details about the EBSD measurements are given in the Supplemental Material (SM)~\cite{SM}.

X-ray absorption spectroscopy (XAS) and RIXS measurements were conducted at the I21-RIXS beamline of the Diamond Light Source, UK \cite{ZhouJSR2022}. All RIXS spectra were collected at temperatures between 22 and 28 K with an effective energy resolution of $\Delta E=37$~meV and a scattering angle of $154^{\circ}$. Incident photons were linearly polarized in either the vertical ($\sigma$) or horizontal ($\pi$) direction with respect to the scattering plane.
The x-ray beam footprint is approximately 2.5~$\upmu$m $\times$ 40~$\upmu$m.

\begin{figure}[tb]
\includegraphics[scale=1]{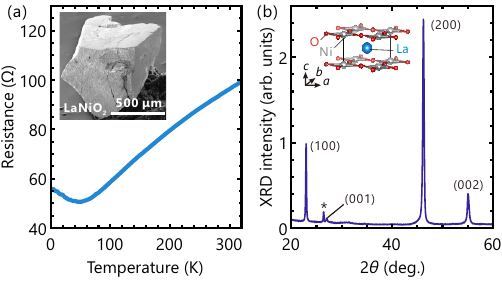}
\caption{(a) Electronic transport of a LaNiO$_{2}$ crystal. 
(Inset) A side-view SEM-SE image of the polished surface of the LaNiO$_{2}$ crystal used in the RIXS experiment. 
(b) XRD pattern acquired from the polished surface of a LaNiO$_{2}$ crystal. The Bragg peaks are indexed and the inset shows the tetragonal $P4/mmm$ unit cell of LaNiO$_{2}$. The  asterisk symbol indicates an artifact originating from a carbon-based paste used to mount the LaNiO$_{2}$ crystal on the sample holder. 
}
\label{fig:resistivity_XRD}
\end{figure}

\begin{figure}[tb]
\includegraphics[scale=1]{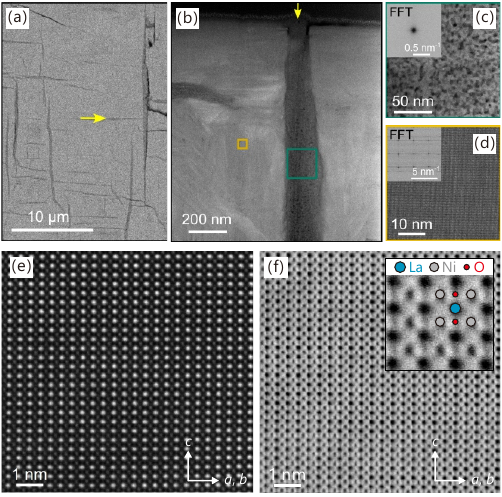}
\caption{(a) Top-view SEM-SE image of the polished surface of a LaNiO$_{2}$ crystal. (b) Cross-sectional view of the crystal in a low-magnification STEM-HAADF image. The yellow arrow marks the same dark line as the arrow in panel (a). [(c),(d)] Magnified view of the area indicated by the green (orange) box in panel (b), which consists of noncrystalline (crystalline) LaNiO$_{2}$. 
(Inset) The fast Fourier transformation. (e) High-magnification STEM HAADF and (f) STEM ABF images acquired within a single domain of a LaNiO$_{2}$ crystal. The inset in (f) zooms into the STEM ABF image, revealing the ideal infinite-layer structure without visible traces of apical oxygen (above and below the Ni atoms) remaining from the perovskite LaNiO$_{3}$ structure.
}
\label{fig:STEM_disorder}
\end{figure}

\section{Results} 

\subsection{Properties of LaNiO$_{2}$ bulk crystals}
\begin{figure}[tb]
\includegraphics[scale=1]{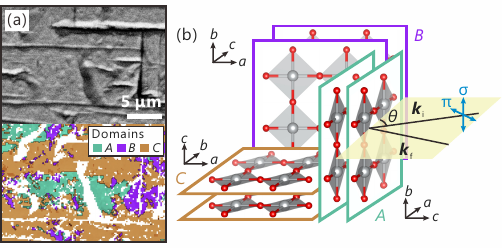}
\caption{(a) SEM-SE image (top panel) and EBSD map (bottom panel) of the same surface region. The color code (green, purple, and brown) indicates the three twin domains ($A$, $B$, and $C$) of LaNiO$_{2}$ with the tetragonal $P4/mmm$ unit cell. White areas correspond to pixels where the EBSD pattern could not be indexed (see the SM~\cite{SM}). 
(b) Schematic of the LaNiO$_{2}$ crystal structure with three twin domains, along with the scattering geometry of the RIXS experiment. The $a$, $b$, and $c$ axes of each domain are indicated, and the square-lattice coordination of Ni (gray) and O (red) is shown for two NiO$_{2}$ planes in each domain, while La ions between the planes are omitted for clarity. The incident photons ($\mathbf{k}_{\rm i}$) are linearly polarized either parallel ($\pi$-pol.) or perpendicular ($\sigma$-pol.) to the scattering plane (yellow). A polarization analysis of the the outgoing photons ($\mathbf{k}_{\rm f}$) was not performed. The incident angle $\theta$ is defined as the angle between $\mathbf{k}_{\rm i}$ and the sample surface.
}
\label{fig:Xtal_domains}
\end{figure}

In a previous study, millimeter-sized LaNiO$_{2}$ crystals with randomly oriented surfaces were obtained via the topotactic reduction of LaNiO$_{3}$ single crystals, utilizing a direct-contact method with CaH$_{2}$ as the reducing agent~\cite{PuphalPRM2023}. Here we employ a similar topotactic method to obtain LaNiO$_{2}$ bulk crystals with crystallographically oriented surfaces (for details see  Sec.~\ref{Methods}). In addition, we polish the crystals to expose fresh surfaces with the LaNiO$_{2}$ phase, which were not affected by the topotactic reduction atmosphere. Figure~\ref{fig:resistivity_XRD}(a) shows the electronic transport of a LaNiO$_2$ crystal measured in a four-point probe geometry on a polished surface [see inset in Fig.~\ref{fig:resistivity_XRD}(a)]. The transport behavior is metallic with a subtle upturn below 50~K, which is consistent with that reported for high-quality LaNiO$_2$ thin films~\cite{Hsu2022}. In contrast, the onset of the upturn in LaNiO$_{2}$ crystals with randomly oriented and unpolished surfaces was approximately 100~K~\cite{PuphalPRM2023}. 

In the x-ray diffraction (XRD) pattern from an oriented and polished surface we observe the characteristic Bragg peaks of LaNiO$_{2}$ in the $P4/mmm$ space group, while no additional peaks due to impurities or insufficiently reduced phases are detected [Fig.~\ref{fig:resistivity_XRD}(b)]. 
However, the simultaneous presence of $(H, 0, 0)$ and $(0, 0, L)$-type Bragg peaks suggests that the LaNiO$_{2}$ crystals contain three twin domains of the tetragonal $P4/mmm$ unit cell. The emergence of these three domains is likely a consequence of the pseudocubic symmetry of the perovskite LaNiO$_{3}$ phase, which lacks a unique preferential direction for oxygen deintercalation during the topotactic reduction process.

A closer inspection of the LaNiO$_{2}$ crystals reveals a distinctive texture on the polished surfaces, characterized by parallel or perpendicular dark lines [Fig.~\ref{fig:STEM_disorder}(a)] that mostly align with the directions of the crystal edges. This texture is both visible in secondary electron (SE) images acquired with a scanning electron microscope (SEM) [Fig.~\ref{fig:STEM_disorder}(a)] and optical microscopy images (see the SM~\cite{SM}). A low-magnification STEM-HAADF image showing a cross-sectional view across a dark line and its vicinity is presented in Fig.~\ref{fig:STEM_disorder}(b). A zoom into the dark area [Fig.~\ref{fig:STEM_disorder}(c)] indicates a porous structure in this region, and the fast Fourier transformation (FFT) of the image suggests that crystalline order is absent [inset in Fig.~\ref{fig:STEM_disorder}(c)]. In contrast, a magnified view of a representative area surrounding the dark lines [Fig.~\ref{fig:STEM_disorder}(d)] reveals a well-ordered crystal lattice. The corresponding FFT image exhibits a pattern that is consistent with the IL structure~\cite{PuphalSciAdv2021}. Further magnification of a crystalline region is provided by high-resolution STEM-HAADF [Fig.~\ref{fig:STEM_disorder}(e)] and ABF imaging [Fig.~\ref{fig:STEM_disorder}(f)], confirming high crystalline quality on a local scale, without any stacking faults or secondary phase inclusions in the lattice. Moreover, the inset in Fig.~\ref{fig:STEM_disorder}(f) illustrates the absence of residual apical oxygen from the LaNiO$_{3}$ phase, indicating that the topotactic transformation to the IL phase is complete.

Notably, the texture featuring dark lines is not present in our unreduced LaNiO$_{3}$ crystals. This suggests that the disordered regions develop during the topotactic reduction, potentially acting as channels that facilitate the oxygen transport from the crystal's interior to its surface. Similar disordered areas separating highly crystalline regions were also identified in a previous STEM study on topotactically reduced Pr$_{0.92}$Ca$_{0.08}$NiO$_{2.75}$ crystals~\cite{Wu2023}. Elucidating the nature of these disordered regions and their role in the topotactic transformation process presents a compelling subject for future STEM investigations.

To further characterize our LaNiO$_{2}$ crystals, we map the distribution of twin domains across the surface using EBSD. The top panel of Fig.~\ref{fig:Xtal_domains}(a) displays an SEM-SE image of a local area on the crystal surface after mechanical and low-angle Ar ion polishing, which is necessary to prepare the surface for EBSD measurements. The bottom panel of Fig.~\ref{fig:Xtal_domains}(a) shows the acquired EBSD map from the same surface area, where we identify three twin domains of the tetragonal space group $P4/mmm$ of LaNiO$_{2}$, color coded in green, purple, and brown. We denote these domains as $A$, $B$, and $C$ in the following. Specifically, the EBSD map reveals extended regions of single-domain character, with domain $C$ predominating. Around the center and the top left corner of the map, significant areas of domain $A$ are observed, interspersed with smaller islands of domain $B$. A mapping of other areas on the crystal surface (see the SM~\cite{SM}), however, reveals varying domain distributions, including regions where domain $A$ or $B$ are dominant. This variation in domain population across areas and length scales of tens of micrometers has important implications for the RIXS experiment, considering that the x-ray beam's footprint on the sample surface is approximately 2.5~$\upmu$m $\times$ 40~$\upmu$m, which is of comparable size. This aspect will be elucidated later in the text.

Interestingly, we find that the nonindexed pixels (white color) in the EBSD maps mostly coincide with dark lines in the associated SE-SEM images, which according to the analysis in Fig.~\ref{fig:STEM_disorder} correspond to disordered regions. The orientation of the twin domains is often identical on both sides of streaks of nonindexed pixels, hence suggesting that these regions are not corresponding to twin domain boundaries. Instead, the boundaries between domains [see different colors in Fig.~\ref{fig:Xtal_domains}(a)], exhibit a more complex structure varying on finer length scales, which also presents a subject for future STEM investigations.

\subsection{RIXS experiment}

The scattering geometry of the RIXS experiment together with the orientation of the three twin domains is depicted in Fig.~\ref{fig:Xtal_domains}(b). We denote the momentum transfer  $(H,K,L)$ in reciprocal lattice units $(2\pi/a,2\pi/b,2\pi/c)$, with the lattice parameters $a,b = 3.9642(3)$~{\AA}, and $c = 3.3561(3)$~{\AA} determined from the powder XRD refinement of a crushed LaNiO$_{2}$ crystal (see the SM~\cite{SM}). For simplicity, we mostly focus on domain $A$ in the following, where the scattering plane is parallel to the crystallographic $a$-$c$ plane [see Fig.~\ref{fig:Xtal_domains}(b)]. 

\subsection{Orbital excitations}
\begin{figure}[tbp]
\includegraphics[scale=1]{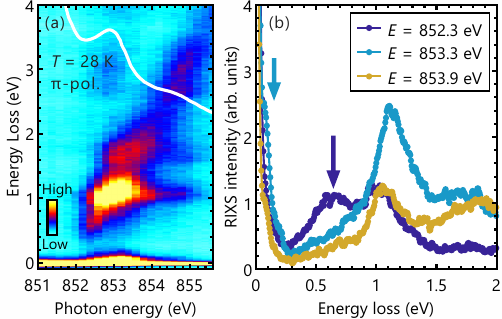}
\caption{(a) RIXS intensity map of the LaNiO$_{2}$ crystal for incident photon energies varied across the Ni $L_{3}$ edge, taken at $\theta=51.2^{\circ}$. The corresponding XAS measured in total electron yield mode is superimposed as the solid white line. The increase of the XAS signal towards low incident energies is due to the tail of the strong La $M_4$-line centered below 851 eV. (b) RIXS spectra for three selected incident photon energies. Arrows at $\sim$0.6 and $\sim$0.15~eV indicate characteristic low energy excitations, as described in the text.}
\label{fig:Emap}
\end{figure}

Figure~\ref{fig:Emap}(a) shows the RIXS spectra of the polished LaNiO$_{2}$ crystal, with the incident photon energy varied across the Ni $L_{3}$-edge. The major spectral weight is centered around 1.2 eV energy loss for incident energies between $E_i = 852.4$ and 853.4 eV, which roughly coincides with the energy range of the main Ni $L_{3}$-edge peak in the XAS [see solid white line in Fig.~\ref{fig:Emap}(a)]. In addition, a sequence of excitations at higher energy losses can be observed in the RIXS intensity map, along with a fluorescence emission feature that streaks across the map up to the highest energies. Overall, the RIXS features in Fig.~\ref{fig:Emap}(a) are closely reminiscent of those reported for LaNiO$_{2}$ thin films \cite{HeptingNatMat2020}, attributable to $dd$ excitations that emerge roughly between 1 and 3 eV energy loss. Subtle differences between the relative intensities of individual $dd$ excitation peaks are likely associated with the presence of three tetragonal twin domains in our LaNiO$_{2}$ crystals, whereas the RIXS map in Ref.~\cite{HeptingNatMat2020} was obtained from a single-domain LaNiO$_{2}$ film. 

Importantly, a close inspection of the region around 0.6 eV energy loss in Fig.~\ref{fig:Emap}(a) reveals the presence of spectral intensity for incident energies around $E_i = 852.3$ eV. In prior thin film studies, this feature has been identified as a hallmark of the infinite-layer nickelates, originating from hybridized La-$5d$ and Ni-$3d$ orbital states \cite{HeptingNatMat2020,RossiPRB2020}. The 0.6 eV feature becomes better discernible when the RIXS spectrum for $E_i = 852.3$ eV is plotted separately [see dark blue arrow in Fig.~\ref{fig:Emap}(b)]. In addition, the RIXS spectra for $E_i = 853.3$ and 853.9 eV are displayed in Fig.~\ref{fig:Emap}(b). While  spectral weight in the latter spectrum shifts towards energy losses above 1.5 eV, we observe a relatively sharp accumulation of spectral intensity below 0.3 eV in the former spectrum (see light-blue arrow). The maximum of this spectral weight is situated around $\sim$0.15~eV. This energy range is above that of phonons, but it is compatible with the bandwidth of spin excitations reported in previous thin-film studies of infinite-layer nickelates~\cite{LuScience2021,RossiNatPhys2022,KriegerPRL2022,Rossi2023}. 

\subsection{Spin excitations}
\begin{figure}[tb]
\includegraphics[scale=1]{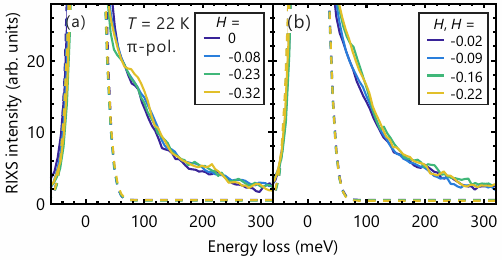}
\caption{(a) RIXS spectra of the LaNiO$_{2}$ crystal for varied momenta along the $(H,0)$ direction, acquired with $\pi$-polarized photons. The dashed curves are Gaussian fits of the elastic line centered at zero-energy loss, using an energy resolution of 37 meV. Substantial spectral weight extends up to $\sim$300~meV, but shows essentially no dispersion as a function of momentum transfer. (b) RIXS spectra along the $(H,H)$ direction.
}
\label{fig:Qmap_LH}
\end{figure}

To elucidate the characteristics of the possible spin excitations, we investigate the momentum-dependence of this spectral weight (Fig.~\ref{fig:Qmap_LH}). To account for photon absorption effects in the crystal for different incident angles, we perform a self-absorption correction of the RIXS data~\cite{RobartsPRB2021,MinolaPRL2015}. Details about the employed procedure can be found in the SM~\cite{SM}. 
Notably, the RIXS spectra along the $(H,0)$ [Fig.~\ref{fig:Qmap_LH}(a)] and $(H,H)$ [Fig.~\ref{fig:Qmap_LH}(b)] directions reveal that the spectral weight is broadly distributed and extends up to $\sim$300~meV for all measured momenta. This is in contrast to the highly dispersive spin wave signal in the RIXS spectra of the nickelate thin films~\cite{LuScience2021,RossiNatPhys2022,KriegerPRL2022}, although the maximum bandwidth across the Brillouin zone is comparable. A subtle increase in spectral intensity is discernible towards higher momenta along both directions, mostly within the energy  range below 100~meV, which is close to the energy regime of phonons. In particular, a prominent phonon peak centered around 60 meV has been reported in Ni $L_3$-edge RIXS experiments on LaNiO$_{2}$ thin films~\cite{RossiNatPhys2022}. Nonetheless, when comparing the RIXS spectra for representative momenta along the $(H,0)$ direction [Fig.~\ref{fig:Qmap_LH}(a)] it appears that the most significant change occurs between 80 and 100 meV. Hence, it is plausible that the broad distribution of low-energy spectral weight in Figs.~\ref{fig:Qmap_LH}(a) and \ref{fig:Qmap_LH}(b) contains contributions from elastic scattering mostly below 40 meV, a phonon around 60 meV, and spin excitations or other contributions at higher energies up to 300 meV.

A difference to single-domain thin films ~\cite{LuScience2021,RossiNatPhys2022,KriegerPRL2022} is that the RIXS signal in our experiment contains the superposition of the three twin domains. 
To test whether this superposition is responsible for the observed ``washed-out'' character of the spin excitations in Figs.~\ref{fig:Qmap_LH}(a) and \ref{fig:Qmap_LH}(b), we simulate spin-wave spectra based on the linear spin-wave theory for the simultaneous presence of three orthogonal crystallographic orientations. 
Specifically, we employ a  $J_1$-$J_2$ model for a square-lattice Heisenberg antiferromagnet, which was also used to model the dispersion measured from an IL nickelate film~\cite{LuScience2021}.
As an approximation, we assume that the RIXS intensity is proportional to the dynamical structure factor of the spin-wave excitations~\cite{AmentRevModPhys2011,JiaNatComm2014}. 
The magnetic Hamiltonian is formulated as follows:
\begin{equation}
{\cal H}=J_{1}\sum_{i,j}\hat{\mathbf{S}}_{i}\cdot\hat{\mathbf{S}}_{j}
+J_{2}\sum_{i,k}\hat{\mathbf{S}}_{i}\cdot\hat{\mathbf{S}}_{k},
\end{equation}
where $\hat{\mathbf{S}}_{i}$ is a spin operator with spin $S=1/2$ at a site $i$.
$J_{1}$ and $J_{2}$ are the nearest and next-nearest-neighbor exchange interactions in the $ab$ plane, respectively. 
We adopt the reported values $J_{1}=63.6$ and $J_{2}=-10.3$~meV from a NdNiO$_{2}$ thin film study~\cite{LuScience2021}, where the spin waves observed in the RIXS spectra exhibited close similarities to those in square-lattice cuprates. 
Accordingly, we define a collinear magnetic structure with an antiferromagnetic ordering vector $\mathbf{Q} = (1/2,1/2)$~\cite{ColdeaPRL2001,ChristensenPNAS2007}.
Using the linear spin-wave theory, the dispersion relation is given by
\begin{equation}
    \hbar\omega_{\mathbf{q}} = 2Z_{c}\sqrt{A_{\bfq}^{2}-B_{\bfq}^{2}},
\end{equation}
where
\begin{eqnarray}
    A_{\bfq} &=& J_{1}-J_{2}\left\{1-\cos{(2\pi H)}\cos{(2\pi K)}\right\} \\
    B_{\bfq} &=& J_{1}\left\{\cos{(2\pi H)}+\cos{(2\pi K)}\right\}/2. 
\end{eqnarray}
The renormalization factor $Z_{c}$ is fixed at 1.18~\cite{SinghPRB1989}, which is expected for a two-dimensional $S=1/2$ square-lattice antiferromagnet.
Further, the inelastic part of the dynamical structure factor is given by:
\begin{equation}
S(\bfq,\omega) \propto \sqrt{\frac{A_{\bfq}-B_{\bfq}}{A_{\bfq}+B_{\bfq}}}
\delta\left(\omega - \omega_{\bfq}\right).
\end{equation}

In analogy to paramagnons in cuprates, which emerge in the absence of long-range order, also the spin excitations in IL nickelates feature a large linewidth~\cite{LuScience2021}. 
Hence, we convolute our calculated spin spectra with a damped harmonic oscillator function. 
We employ a damping factor of 74~meV, in accord with the average of the momentum-dependent damping factors in Ref.~\cite{LuScience2021}.
Furthermore, our damped spectra are convoluted with Gaussian functions along the energy and momentum directions according to the RIXS experimental energy resolution of 37~meV and the $q$ resolution of 0.09~\AA$^{-1}$, respectively The $q$ resolution is estimated from the mosaicity of the LaNiO$_{2}$ crystal according to the full width at half maximum of the $(001)$ Bragg peak.

\begin{figure}[tb]
\includegraphics[scale=1]{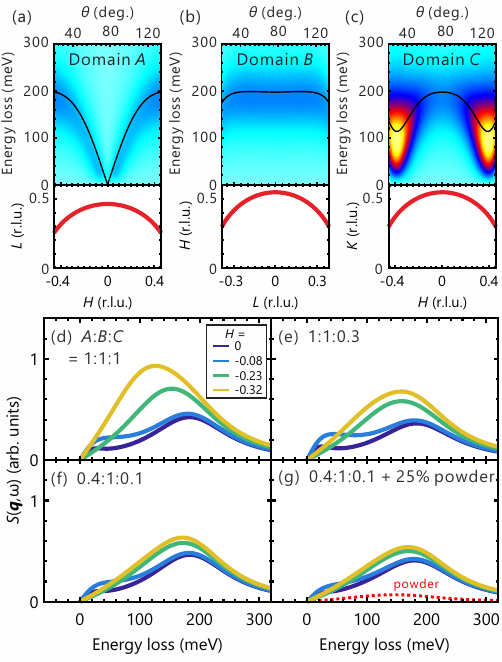}
\caption{Spin wave simulation for a system with three twin domains and the scattering geometry used in the RIXS experiment, with varying incident angle $\theta$, fixed scattering angle at 154$^{\circ}$, and $E_{i}=853$~eV. (a)--(c) False-color plots (top panels) of the simulated spin-wave spectra of the twin domains $A$, $B$, and $C$, respectively. The intensity scale corresponds to the dynamical structure factor $S(\mathbf{q},\omega)$ and the solid black lines are the dispersion relations. The solid red lines (bottom panels) indicate the corresponding momentum transfer for each domain within the $H-L$, $L-H$, and $K-H$ planes, respectively. (d)--(g) Superposition of $S(\mathbf{q},\omega)$ from the three domains assuming different populations of the domains $A$, $B$, and $C$. For better comparability to the RIXS data in Fig.~\ref{fig:Qmap_LH}(a), the momenta in panel (d) are indexed according to domain $A$ with variation along the $(H, 0)$ direction, although the calculated $S(\mathbf{q},\omega)$ contains the contributions from all three domains. The domain population ratios in panels (d)--(f) are $1 : 1 : 1$, $1 : 1 : 0.3$, and $0.4 : 1 : 0.1$, respectively. In panel (f), an almost non-dispersive behavior of the spin wave is reproduced, which can be further enhanced by taking into account a 25{\%} volume fraction with randomly oriented grains or disordered crystal structure (powder).
}
\label{fig:SW_1Dcut}
\end{figure}

Figures~\ref{fig:SW_1Dcut}(a)--\ref{fig:SW_1Dcut}(c) present the simulated spin wave spectra for three orthogonal crystal orientations, each contributing a distinct spin-wave signal to the RIXS spectrum. 
The three orientations represent the twin domains illustrated in Fig.~\ref{fig:Xtal_domains}(b). 
In our RIXS experiment, the momentum transfer is varied by changing the incident angle $\theta$ [see Fig.~\ref{fig:Xtal_domains}(b)], while keeping the scattering angle fixed at $154^{\circ}$. 
As a consequence, a change of the in-plane momentum transfer also involves a change of the out-of-plane momentum. For a single-domain sample, such as an IL nickelate thin film, the change in the out-of-plane momentum is usually neglected since the dispersion of the spin waves along the $L$ direction is flat for a quasi-two-dimensional square lattice antiferromagnet. 
However, for the simultaneous presence of three twin domains the situation is more involved and hence requires the separate consideration of each domain [Figs.\ref{fig:SW_1Dcut}(a)--\ref{fig:SW_1Dcut}(c)]. The corresponding trajectories in reciprocal space for the three domains when the incident angle is varied from $\theta = 20^{\circ}$ to 134$^{\circ}$ are shown in the bottom panels in Figs.~\ref{fig:SW_1Dcut}(a)--\ref{fig:SW_1Dcut}(c).

Assuming an equal population $(1:1:1)$ of the $A$, $B$, and $C$ twin domains, we obtain a composite spectrum shown in Fig.~\ref{fig:SW_1Dcut}(d). The simulated peak positions disperse significantly more weakly as a function of $H$ than the spin waves of single-domain thin films ~\cite{LuScience2021}. 
However, a strong increase in intensity for the highest momenta $H$ in Fig.~\ref{fig:SW_1Dcut}(d) is not present in our RIXS data in Fig.~\ref{fig:Qmap_LH}(a). 
Considering that the typical lateral dimensions of the domains in the LaNiO$_{2}$ crystal are between a few and several tens of micrometers (see Fig.~\ref{fig:Xtal_domains}(a) and the SM~\cite{SM}) while the footprint of the x-ray beam on the sample surface is 2.5 $\upmu$m in the vertical direction and more than 40 $\upmu$m in the horizontal direction (depending on the incident angle $\theta$), it is plausible to assume that the population of the three domains in the probed sample volume was not equal. 
Accordingly, we next simulate unequal domain populations of $1:1:0.3$ and $0.4:1:0.1$ in Figs.~\ref{fig:SW_1Dcut}(e) and \ref{fig:SW_1Dcut}(f), respectively. 
A notable trend in these simulations is a further reduction of the dispersive character and intensity enhancement, aligning more closely with the RIXS data presented in Fig~\ref{fig:Qmap_LH}(a). 

As a final step, we take into account the presence of the textures with dark lines in our LaNiO$_{2}$ crystals [Figs.~\ref{fig:STEM_disorder}(a)--\ref{fig:STEM_disorder}(c)], indicative of regions with a disordered crystal structure. From Fig.~\ref{fig:STEM_disorder} and optical microscope images (see the SM~\cite{SM}) we estimate that these regions contribute at most 25{\%} of the the sample's volume fraction, and we incorporate this sample characteristic as powder-averaged spin-wave spectra in our simulation. Although the simulation result   [Fig.~\ref{fig:SW_1Dcut}(g)] does still not exactly match our RIXS data, the trend is notably consistent. Considering that we used the damping and magnetic exchange parameters from previous thin film studies as sole input for our simulation, it suggests that the magnetic correlations within a single domain of our crystals closely resemble those in single-domain films. The remaining discrepancy observed in Fig.~\ref{fig:SW_1Dcut}(g) might be due to increased damping of the spin waves in the crystals or other factors, highlighting the need for further investigation.

\subsection{Charge order}
Next, we examine whether the charge order phenomenology reported for IL nickelate thin films similarly occurs in crystals. Specifically, a number of previous RIXS studies on thin films performed under comparable experimental conditions, reported the occurrence of a CDW signal in the form of enhanced quasielastic scattering around $H=\pm 1/3$~\cite{RossiNatPhys2022,KriegerPRL2022,TamNatMat2022}.

\begin{figure}[tbp]
\includegraphics[scale=1]{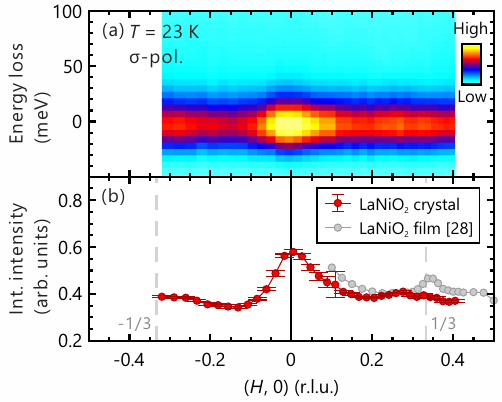}
\caption{(a) RIXS intensity map of the LaNiO$_{2}$ crystal for momenta varied along the $(\pm H,0)$ direction, acquired with $\sigma$-polarized incident photons. (b) Integrated intensity of the quasielastic scattering along the $(\pm H,0)$ direction. The integration range is $-50$~meV $\leq E \leq 50$~meV energy loss. Error bars are determined by the standard deviation of six RIXS spectra measured at each momentum. In addition, the quasielastic intensity from a RIXS experiment on a LaNiO$_2$ thin film in Ref.~\cite{RossiNatPhys2022} is superimposed as gray symbols, exhibiting a distinct CDW peak around $H = 0.344$. The RIXS intensities of the crystal and the film are both normalized to the integrated intensity of the $dd$ excitations (see the SM~\cite{SM}). In addition, a vertical offset is added to the film intensity to match the baseline of the crystal data. }
\label{fig:Qmap_LV}
\end{figure}

Figure~\ref{fig:Qmap_LV}(a) shows a RIXS intensity map of the region of quasielastic scattering of the LaNiO$_2$ crystal for different momenta along the $(H,0)$ direction, utilizing $\sigma$-polarized incident photons to enhance the charge signal. Note that these RIXS spectra were obtained from a different surface spot on the LaNiO$_2$ crystal compared to those in Fig.~\ref{fig:Qmap_LH}. Hence, the underlying domain population is likely distinct, and for simplicity the momentum transfer in Fig.~\ref{fig:Qmap_LV} is indexed with respect to domain $A$. The self-absorption correction \cite{RobartsPRB2021,MinolaPRL2015} was applied to all spectra in Fig.~\ref{fig:Qmap_LV}(a). 

Remarkably, we find that the data from our LaNiO$_2$ crystal lack clear charge ordering signatures around $H=\pm 1/3$  [Fig.~\ref{fig:Qmap_LV}(a)]. This absence of an enhanced quasielastic signal is  corroborated when examining the integrated RIXS intensity between $-50$~meV $\leq E \leq 50$~meV energy loss [red data points in Fig.~\ref{fig:Qmap_LV}(b)]. 
For a direct contrast, we also plot LaNiO$_{2}$ film data from Ref.~\cite{RossiNatPhys2022} in Fig.~\ref{fig:Qmap_LV}(b).
Note that we have normalized both our RIXS data and those of Ref.~\cite{RossiNatPhys2022} 
 by the integrated intensity of the respective $dd$ excitations to ensure direct comparability.

While the pronounced $H=\pm 1/3$ peaks of Ref.~\cite{RossiNatPhys2022} are clearly absent in our data, we observe a broad peak-like feature around $H\sim0.27$, which is subtle but exceeds the range of our experimental error bars. Yet, for momentum transfer in the opposite direction, a corresponding peak feature at $H\sim-0.27$ is not visible. Instead, the quasielastic scattering intensity in this direction increases continuously from $H\sim-0.15$ to $-0.3$. To verify that these behaviors are not associated with CDW signals, we carried out additional RIXS measurements on other locations on the surface of the same crystal as well as on two additional crystals (see the SM~\cite{SM}), which did not reproduce the peak at $H\pm0.27$. In consequence, we attribute such deviations from the baseline intensity of quasielastic scattering to artifacts resulting from the partially uneven surface topography of our LaNiO$_2$ crystals (see inset in Fig.~\ref{fig:resistivity_XRD}(a) and the SM~\cite{SM}).

Nonetheless, due to the presence of three twin domains in our LaNiO$_2$ crystals, we cannot fully rule out that a CDW signal has remained undetected in our experiment. For instance, one scenario could involve an unfavorable domain distribution within the surface area probed by RIXS in Fig.~\ref{fig:Qmap_LV} where the domains $A$ and $B$ would be essentially unpopulated, while for the dominating domain $C$ our applied momentum transfer range would not cover the nominal $\mathbf{q}=(\pm 1/3,0)$ ordering vector. However, this scenario seems implausible, as our EBSD analysis of various crystal surface areas (see the SM~\cite{SM}) revealed that typically at least two twin domains exhibit significant population within an area corresponding to the x-ray beam footprint of approximately 2.5~$\upmu$m $\times$ 40~$\upmu$m. Moreover, our search for a CDW signal encompassed different spots on multiple LaNiO$_2$ crystals (see the SM~\cite{SM}), making it even less likely that we consistently encountered unfavorable domain distributions. 

The absence of the $\mathbf{q}=(\pm 1/3,0)$ CDW peaks in our LaNiO$_2$ sample is particularly noteworthy when considered in conjunction with the STEM characterization of our crystal [Fig.~\ref{fig:STEM_disorder}(f)]. Specifically, the absence of any traces of residual apical oxygen indicates that the ideal IL structure is realized in our crystals. This concomitant absence of apical oxygen and a CDW signal is compatible with the notion that excess oxygen arranged in patterns characterized by a $\mathbf{q}=(\pm 1/3,0)$ ordering vector is a prerequisite for the emergence of charge order in nickelates \cite{RajiSmall2023,Parzyck2024}. 

\section{Conclusions} 
In summary, our results have several important implications for the ongoing debate about  spin and charge instabilities in the material class of IL nickelates. In particular, our observation of spin excitations with a bandwidth of more than 200~meV akin to those reported for thin films ~\cite{LuScience2021,KriegerPRL2022,RossiNatPhys2022} suggests that the associated spin dynamics are an ubiquitous feature of the pure IL phase. Nevertheless, the absence of a discernible spin-wave dispersion in our data, presumably due to a distribution of twin domains within the probed region on the crystal surface, calls for further investigations. In particular, the synthesis of single-domain LaNiO$_{2}$ crystals becomes imperative for future experimental work, potentially achievable by applying strain or pressure during the topotactic reduction, akin to the epitaxial strain conditions that yield single-domain IL thin films \cite{Lee2023,SunAdvMat2023,ChowFront2022}.

Furthermore, the lack of a $\mathbf{q}=(\pm 1/3,0)$ CDW signal in our data a suggests that charge order is not intrinsic to the bulk phase of IL nickelates. This notion is compatible with the scenario that a secondary phase of residual apical oxygen ions hosts the CDW~\cite{RajiSmall2023,Parzyck2024}, or that epitaxial strain from a substrate plays a role in the emergence of charge order in IL nickelates \cite{Rossi2023}. In both cases, further investigations are warranted to clarify if analogies between the $\mathbf{q}=(\pm 1/3,0)$ nickelate charge order phenomenology and that in cuprates \cite{GhiringhelliScience2012,ChangNatPhys2012,BlancoPRB2014,LiPNAS2020,Frano2020,ArpaiaJPSJ2021,LeeNatPhys2021,LeeCDW2021} exist. 

Given our focus on the undoped parent compound LaNiO$_{2}$ and axial directions up to wave vectors of $\mathbf{q}=(\pm 0.4,0)$, our results do not rule out the existence of alternative charge ordering patterns in IL nickelates with different vectors and/or an emergence upon hole doping. This highlights the need for future studies on IL nickelate bulk crystals and thin films with assured stoichiometric oxygen content, with a focus either on exploring the phase space suggested by the CDW in cuprates, or undertaken as a comprehensive survey of the entire experimentally accessible reciprocal space and doping parameters.

\begin{acknowledgements}
We thank W.-S.~Lee, J.~Weis, and A.~Krajewska for helpful discussions, and C. Busch, U. Waizmann, and M. Hagel for technical support. We thank J.~Deuschle for FIB lamella preparation. We acknowledge Diamond Light Source for providing the beamtime under the Proposal No. MM33241-1, and are grateful to the technical staff for assistance during the experiment.
\end{acknowledgements}

\bibliography{LaNiO2_RIXS}

\end{document}


\title{Supplemental Material for ``Investigation of spin excitations and charge order in bulk crystals of the infinite-layer nickelate LaNiO$_2$"}

\author{S.~Hayashida}
\email[]{s.hayashida@fkf.mpg.de}
\affiliation{Max-Planck-Institute for Solid State Research, Heisenbergstra$\beta$e 1, 70569 Stuttgart, Germany}
\author{V.~Sundaramurthy}
\affiliation{Max-Planck-Institute for Solid State Research, Heisenbergstra$\beta$e 1, 70569 Stuttgart, Germany}
\author{P.~Puphal}
\affiliation{Max-Planck-Institute for Solid State Research, Heisenbergstra$\beta$e 1, 70569 Stuttgart, Germany}
\author{M.~Garcia-Fernandez}
\affiliation{Diamond Light Source, Harwell Campus, Didcot OX11 0DE, United Kingdom}
\author{Ke-Jin~Zhou}
\affiliation{Diamond Light Source, Harwell Campus, Didcot OX11 0DE, United Kingdom}
\author{B.~Fenk}
\affiliation{Max-Planck-Institute for Solid State Research, Heisenbergstra$\beta$e 1, 70569 Stuttgart, Germany}
\author{M.~Isobe}
\affiliation{Max-Planck-Institute for Solid State Research, Heisenbergstra$\beta$e 1, 70569 Stuttgart, Germany}
\author{M.~Minola}
\affiliation{Max-Planck-Institute for Solid State Research, Heisenbergstra$\beta$e 1, 70569 Stuttgart, Germany}
\author{Y.-M.~Wu}
\affiliation{Max-Planck-Institute for Solid State Research, Heisenbergstra$\beta$e 1, 70569 Stuttgart, Germany}
\author{Y.~E.~Suyolcu}
\affiliation{Max-Planck-Institute for Solid State Research, Heisenbergstra$\beta$e 1, 70569 Stuttgart, Germany}
\author{P.~A.~van~Aken}
\affiliation{Max-Planck-Institute for Solid State Research, Heisenbergstra$\beta$e 1, 70569 Stuttgart, Germany}
\author{B.~Keimer}
\email[]{b.keimer@fkf.mpg.de}
\affiliation{Max-Planck-Institute for Solid State Research, Heisenbergstra$\beta$e 1, 70569 Stuttgart, Germany}
\author{M.~Hepting}
\email[]{hepting@fkf.mpg.de}
\affiliation{Max-Planck-Institute for Solid State Research, Heisenbergstra$\beta$e 1, 70569 Stuttgart, Germany}
 
\date{\today}

\begin{abstract}
\end{abstract}

\maketitle

\section{Characterization of reduced crystals}



\begin{figure}[tbp]
\includegraphics[scale=1]{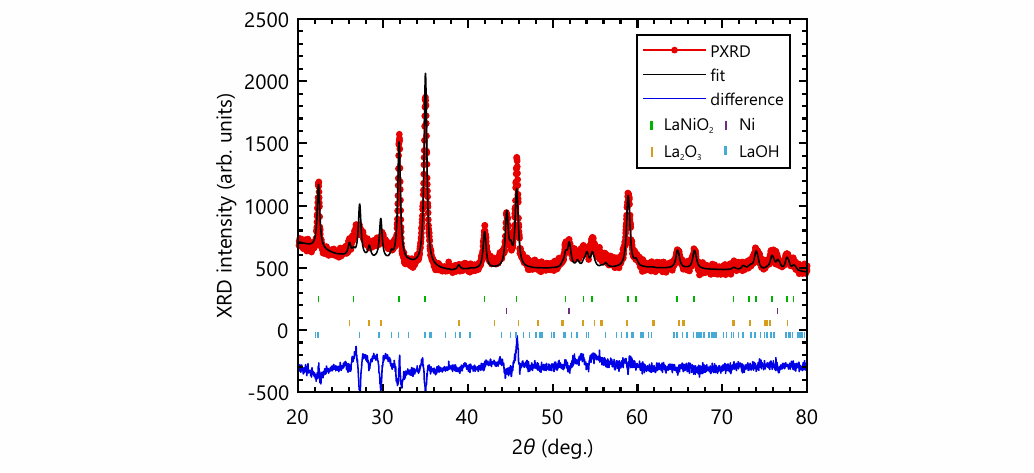}
\caption{PXRD pattern of a pulverized crystal after topotactic reduction. The Rietveld refinement includes various minority phases besides the LaNiO$_{2}$ majority phase, as indicated in the legend.}
\label{fig:XRD}
\end{figure}

To determine the progress of the topotactic reduction in the crystals, we crushed a piece fractured off from a cube-shaped crystal to powder, which was examined with powder x-ray diffraction (PXRD), using a Rigaku Miniflex diffractometer with Cu $K\alpha$ radiation. 
Supplementary Fig.~\ref{fig:XRD} shows the corresponding PXRD pattern, where the main Bragg peaks can be indexed in the tetragonal space group $P4/mmm$ of LaNiO$_2$. 
The refined lattice parameters of LaNiO$_2$ are $a, b = 3.9642(3)$~{\AA}, and $c = 3.3561(3)$~{\AA}. 
Additional PXRD peaks are indexed by elemental Ni, La$_2$O$_3$, and lanthanum oxide hydride LaOH. 
The volume fractions of each phase are calculated approximately as 54{\%} LaNiO$_{2}$, 21{\%} Ni, 16{\%} La$_{2}$O$_{3}$, and 9{\%} LaOH.
These phases were also identified in Ref.~\cite{PuphalPRM2023} for a decomposed LaNiO$_3$ crystal after excessive reduction. 
Notably, our PXRD exhibits no traces of residual LaNiO$_3$ and LaNiO$_{2.5}$ phases. The relatively large volume fraction of decomposed phases in the PXRD in Supplementary Fig.~\ref{fig:XRD} is likely because the investigated fractured piece exhibited a large surface area that was exposed to the topotactic reduction atmosphere. By contrast, we find that XRD investigations of crystal surfaces after mechanical polishing do not show any phases other than LaNiO$_2$ [Fig.~1(b) in the main text], suggesting that only the outer layers of the crystal decompose during the topotactic reduction process. 



A secondary electron (SE) image taken with a Tescan Vega (TS-5130MM) scanning electron microscope (SEM) of the crystal used for the RIXS experiment is shown in Supplementary Fig.~\ref{fig:Xtal}(a). Note that the surface area of the measured crystal [Supplementary Fig.~\ref{fig:Xtal}(b)] is smaller than the original $0.8 \times 0.8$ mm$^2$ since parts of the crystal fractured off during the polishing and sample mounting procedures. The enlarged SEM-SE image in Supplementary Fig.~\ref{fig:Xtal}(c) reveals that the polished surface exhibits a pattern of rectangular structures, separated by either large fractures with voids or smaller cracks. 

\section{Analysis of twin domains}

We used electron energy backscatter diffraction (EBSD) to investigate the microstructural properties of the LaNiO$_{2}$ crystals, such as the twin domain distribution. Since EBSD is a surface-sensitive technique in which the diffraction signals originate from the top few nanometers of the sample's surface, it is crucial to ensure that the surface is free of lattice damage and contamination. Accordingly, after the RIXS experiment, we subjected the LaNiO$_{2}$ crystal to additional mechanical polishing followed by low-voltage, low-angle Ar ion polishing to prepare a surface suitable for EBSD studies. Specifically, the crystal was glued on a conductive plate via Ag-epoxy and then embedded in a two-component epoxy resin (cured for 15 min at $110^\circ$C). The excess epoxy resin above the crystal was removed using 15 $\mu$m SiC sandpaper in a Leica EM-TXP preparation system, and the crystal surface was polished sequentially with 9, 2, and 0.5 $\mu$m diamond lapping films. The Ar ion polishing was performed in a Jeol IB-09010CP Ar Ion Slope Cutter for 50 min at 2.5 kV and 30 $\mu$A, at an inclination angle of $4^\circ$. The sample was continuously rotated to avoid curtaining effects and reduce the formation of ramp-like surface structures around the cracks and grooves on the surface of the crystals. 

Supplementary Figs.~\ref{fig:optical}(a) and \ref{fig:optical}(b) show optical bright-field images of the LaNiO$_{2}$ crystal after the additional mechanical polishing. We note a contrast between the LaNiO$_{2}$ phase and a small admixture of Ni impurities, which appear as bright spots in the optical images. These impurities likely originate from eutectic NiO inclusions in the OFZ grown LaNiO$_{3}$ crystal \cite{PuphalAPL2023}, which were topotactically reduced to elemental Ni. After the Ar ion polishing, the surface topography exhibits excavations with inclinations that mostly coincides with the $4^\circ$ inclination angle of the Ar ion beam on the crystal surface [Supplementary Figs.~\ref{fig:optical}(c)]. 

Next, EBSD is performed on representative regions on the surface of the LaNiO$_{2}$ crystal. Supplementary Figs.~\ref{fig:ebsd1}(a)-\ref{fig:ebsd1}(c) show SEM images of the first investigated region, acquired with different SEM detectors (an Everhart-Thornley-type standard secondary electron detector and 5 diode detectors are grouped around the EBSD scintillator). Several rectangular cracks and excavations occur in this region. Supplementary Fig.~\ref{fig:ebsd1}(d) displays a map of the EBSD band contrast in the same region, representing the quality of the EBSD pattern acquired at individual pixels of the map. While the quality in several areas of the map is high (dark gray pixels), the quality within the cracks and deep excavations ranges from low (light gray pixels) to no diffraction pattern (white pixels). In Supplementary Fig.~\ref{fig:ebsd1}(e), the red color indicates which pixels in the band contrast map yielded an electron diffraction pattern that could be indexed in the $P4/mmm$ unit cell of LaNiO$_{2}$ \cite{Crespin2005}. The indexing rate of LaNiO$_{2}$ among the light gray to dark gray pixels is higher than 90\%, suggesting the presence of a homogeneous LaNiO$_{2}$ phase in the investigated region, in accord with the XRD result of a crystal surface [Fig.~1(b) in the main text]. 
Note that the fact that some light gray pixels are not indexed as LaNiO$_{2}$, as well as the presence of white pixels with no diffraction pattern, does not necessarily imply the presence of a phase other than LaNiO$_{2}$ in these regions. Instead, the uneven surface topography and/or degraded crystallinity of LaNiO$_{2}$ within the cracks and deep excavations may have affected the quality of the corresponding diffraction patterns. Further studies are needed to clarify the composition and material properties within these surface textures.

Supplementary Fig.~\ref{fig:ebsd1}(f) shows the orientation of the $P4/mmm$ unit cell of LaNiO$_{2}$ for each indexed pixel, with the three tetragonal twin domains denoted as $A$, $B$, and $C$ color-coded in green, purple, and brown, respectively.  The orientations of the three domains relative to the sample's surface are illustrated in Supplementary Figs.~\ref{fig:ebsd1}(g)-\ref{fig:ebsd1}(i), accompanied by representative electron diffraction patterns of each respective domain. In Supplementary Fig.~\ref{fig:ebsd1}(f), we observe extended areas of single-domain character (especially of domain $C$), spanning several tens of micrometers, interspersed with smaller islands of other domains and unindexed pixels. Around the center of the investigated region and in the upper left corner of the map, sizable areas of domain $A$ are present. Importantly, the EBSD investigation reveals that individual domains extend across the horizontal and vertical cracks observed in the SEM images in Supplementary Figs.~\ref{fig:ebsd1}(a)-\ref{fig:ebsd1}(c), which appear as (almost) continuous horizontal and vertical lines in the EBSD map in Supplementary Fig.~\ref{fig:ebsd1}(f). This suggests that horizontal and vertical cracks, also visible in Supplementary Fig.~\ref{fig:Xtal}(c) and Figs.~\ref{fig:optical}(a)-\ref{fig:optical}(c), do not correspond to grain boundaries between the three tetragonal twin domains. Instead, the twin boundaries extracted in Supplementary Fig.~\ref{fig:ebsd1}(f) exhibit a more complex, fractal-like nature with finer length scales, the exploration of which will be an interesting topic for future studies.  

Supplementary Fig.~\ref{fig:ebsd2} presents SEM images and EBSD maps of the second investigated surface region of the LaNiO$_{2}$ crystal. Similar to the first region examined in Supplementary Fig.~\ref{fig:ebsd1}, the second region also exhibits rectangular cracks and deep excavations, which could not be indexed in the $P4/mmm$ unit cell of LaNiO$_{2}$ [Supplementary Fig.~\ref{fig:ebsd2}(e)]. Nevertheless, the regions indexed as LaNiO$_{2}$ constitute the three tetragonal twin domains $A$, $B$, and $C$. Notably, domain $C$ is the most extended one in the first examined region [Supplementary Fig.~\ref{fig:ebsd1}(f)], while domain $B$ is most extended in Supplementary Fig.~\ref{fig:ebsd2}(f). 

The results of the investigation of a third surface region are displayed in Supplementary Fig.~\ref{fig:ebsd3}. Another distinct domain distribution is observed, suggesting that the RIXS signal in our experiment, originating from the footprint of the x-ray beam, spanning 2.5 $\mu$m in the vertical and more than 40 $\mu$m in the horizontal direction (depending on the incident angle $\theta$), may contain contributions from non-equal domain distributions.

\section{Self-absorption correction of RIXS spectra}
We apply a self-absorption correction~\cite{RobartsPRB2021,MinolaPRL2015} to the RIXS spectra to account for absorption effects at different incident angles, corresponding to different in-plane momentum transfer. Accordingly, we acquired XAS spectra at $\theta = 21.2^{\circ}$ and $91.2^{\circ}$ for both $\pi$ and $\sigma$ polarizations, which were used as an input for the correction procedure described in Ref.~\cite{RobartsPRB2021}. 
Supplementary Fig.~\ref{fig:self_absorption} shows the raw data together with the corrected RIXS intensity maps, in the context of Fig. 3 of the main text.
All RIXS spectra were acquired with $\pi$-polarized incident photons, while no polarization analysis of the outgoing photons was performed. The corrected spectra for three possible polarization configurations of the incident and outgoing photons are displayed in Supplementary Fig.~\ref{fig:self_absorption}: (i) averaging over the $\pi$-$\sigma$ and $\pi$-$\pi$ configurations [Supplementary Fig.~\ref{fig:self_absorption}(b)], (ii) the $\pi$-$\sigma$ configuration [Supplementary Fig.~\ref{fig:self_absorption}(c)], and (iii) the $\pi$-$\pi$ configuration [Supplementary Fig.~\ref{fig:self_absorption}(d)].
The major effect of the correction for all configurations is a modest attenuation of spectral intensities at high momenta. However, the corrected spectra for all different configurations remain similar to each other, justifying our focus on the averaged $\pi$-$\sigma$ and $\pi$-$\sigma$ configurations for the RIXS spectra presented in Fig.~3 in the main text.

In analogy, we also employed the averaging over the $\sigma$-$\pi$ and $\sigma$-$\sigma$ configurations for the RIXS spectra in the map of Fig.~7 in the main text.

\section{Complementary RIXS spectra and absence of charge order signal}
\begin{figure}[tbp]
\includegraphics[scale=1]{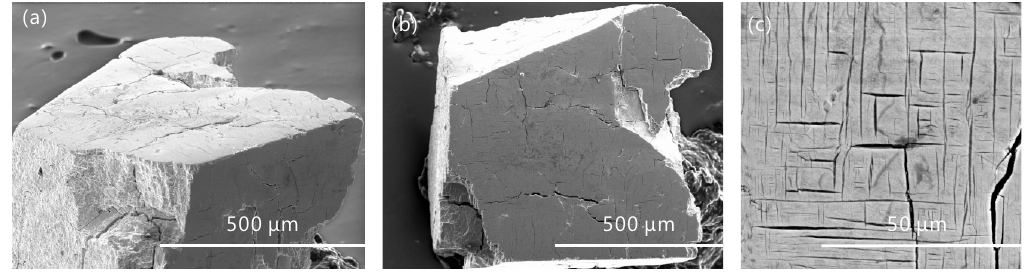}
\caption{(a) Side-view SEM-SE image of the polished surface of the LaNiO$_{2}$ crystal used in the RIXS experiment. (b) Top-view of the polished surface of the same crystal. (c) Zoom into a small region on the surface. The majority of the smaller fractures between the rectangular structures is approximately aligned with either the $a$, $b$, or $c$ direction of the $P4/mmm$ unit cell. 
}
\label{fig:Xtal}
\end{figure}

\begin{figure}[tb]
\includegraphics[width=1\columnwidth]{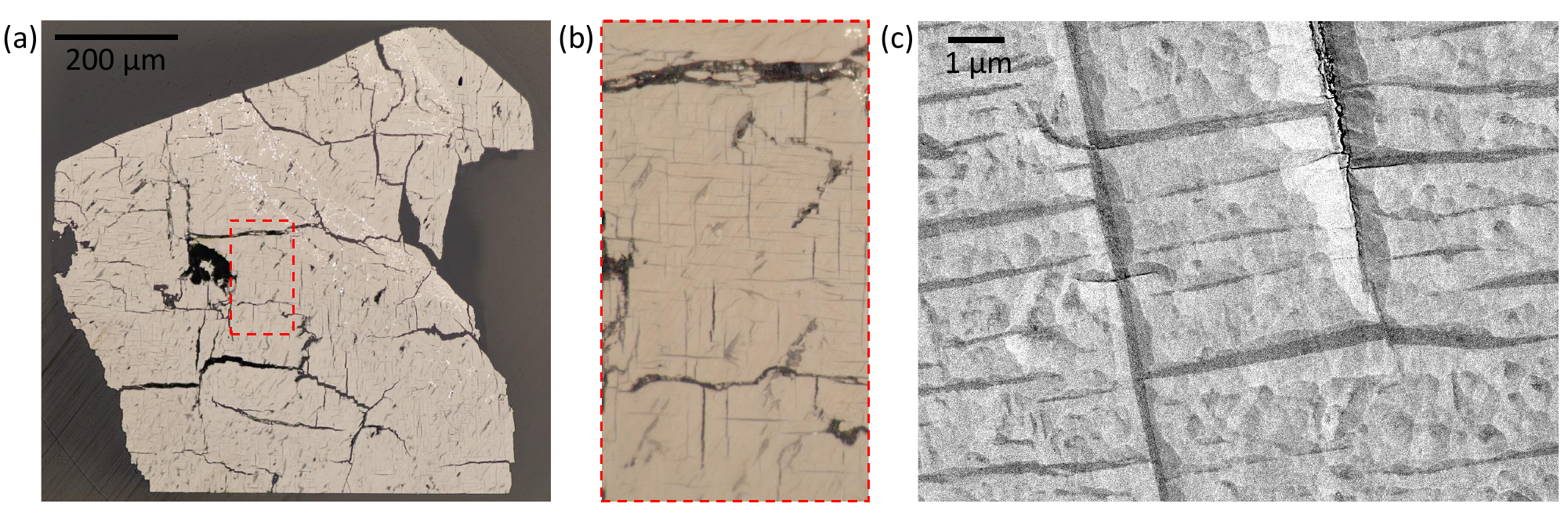}
\caption{(a) Top-view optical bright field image of the LaNiO$_{2}$ crystal after additional mechanical polishing and Ar ion milling. (b) Zoom into the region of the red dashed box in panel (a). (c) SEM SE image on a local scale, revealing a rough topography of the surface due to the Ar ion milling. }
\label{fig:optical}
\end{figure}

\begin{figure}[tb]
\includegraphics[width=1\columnwidth]{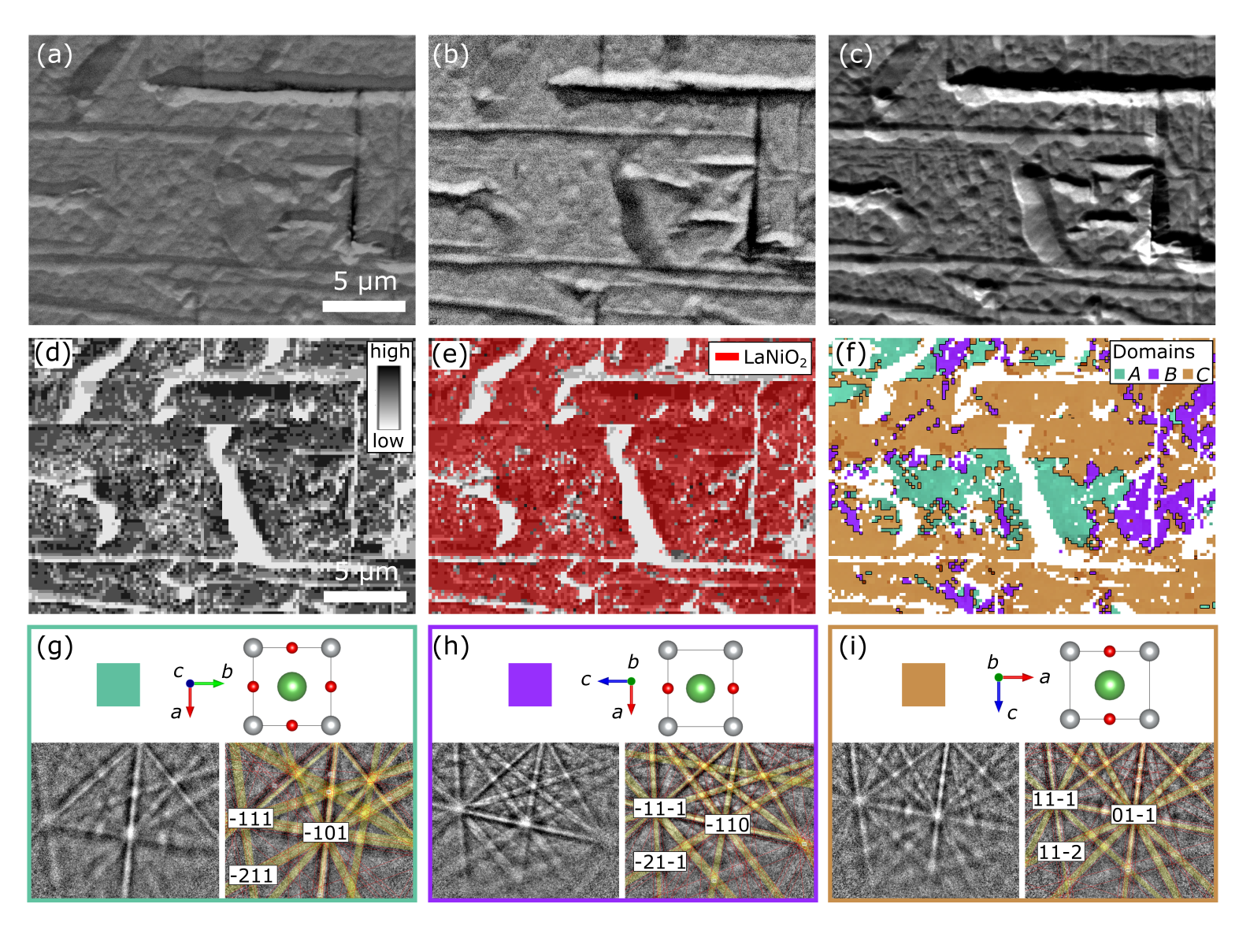}
\caption{(a)-(c) SEM images of the first surface region of the LaNiO$_{2}$ crystal selected for the EBSD investigation. (a) The composite SEM image from all SEM detector inputs, (b) a backscatter detector image highlighting chemical and topographical contrast, and (c) a forward scattering detector image emphasizing topographical contrast. (d)-(f) EBSD mapping of the surface. (d) Band contrast representing the quality of the EBSD pattern at individual pixels of the map, with the gray scale ranging from high quality (dark gray) to low quality (light gray), as well as the absence of an electron diffraction pattern (white). (e) Superimposed band contrast and phase map, with pixels shown in red color where the EBSD pattern could be indexed with the $P4/mmm$ unit cell of LaNiO$_{2}$. (f) Domain map showing for each indexed pixel the crystallographic orientation of the $P4/mmm$ unit cell. The color code green, purple, and brown corresponds to the three twin domains $A$, $B$, and $C$, respectively. Different shades of individual colors in panel (f) mark grains within a domain that exhibit small misorientations of a few degrees. (g)-(i) The $a$, $b$, and $c$ axis directions and $P4/mmm$ unit cell orientation within each domain, projected along the normal direction of the sample surface. The panels (g)-(i) also display representative electron diffraction patterns from each domain, acquired at a 70$^\circ$ tilt angle of the sample. 
The left panel shows the raw data of the Kikuchi lines and the right panel the indexed counterparts with the simulated bands (yellow lines).}    
\label{fig:ebsd1}
\end{figure}

\begin{figure}[tb]
\includegraphics[width=1\columnwidth]{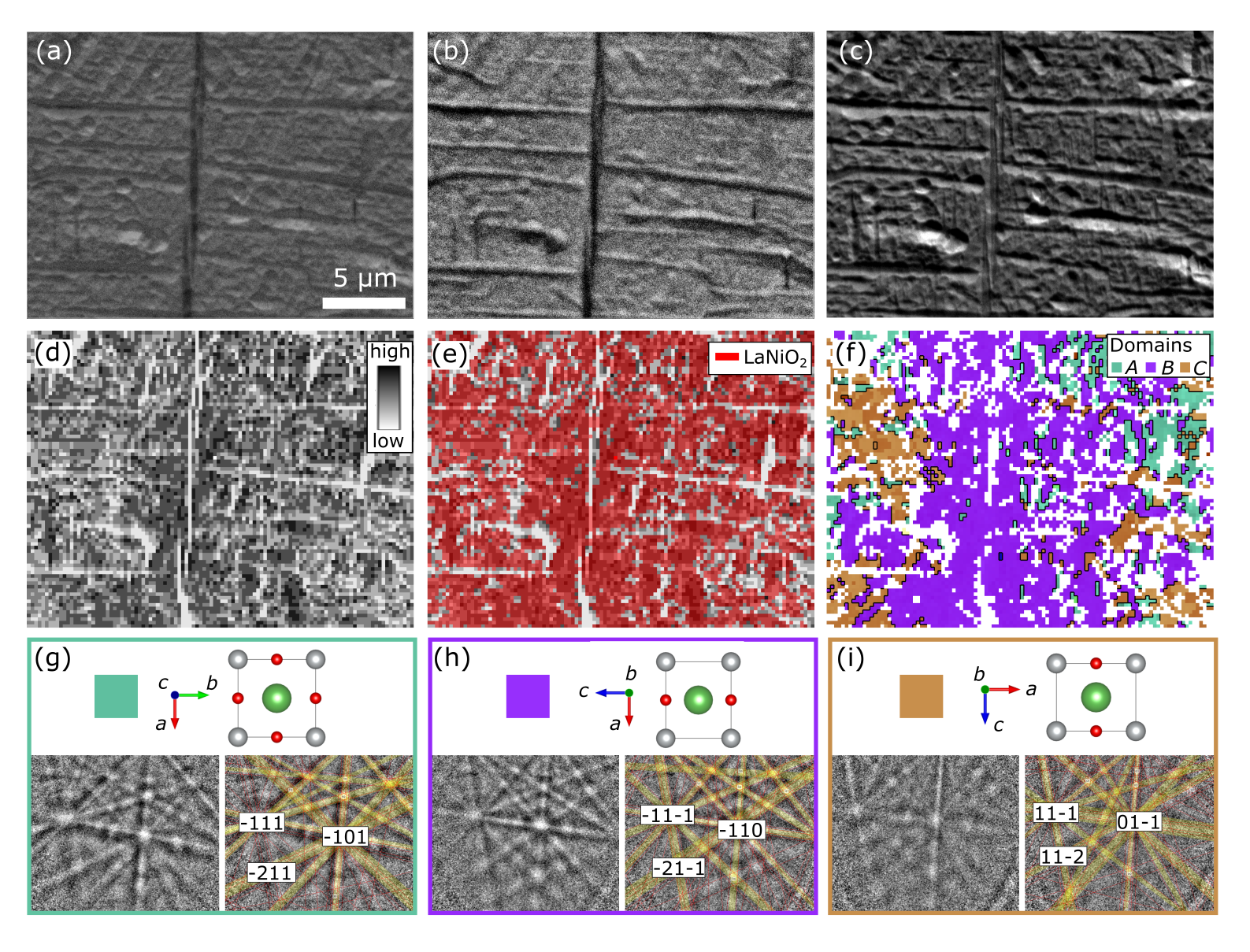}
\caption{SEM images and EBSD maps of the second surface region of the LaNiO$_{2}$ crystal  selected for the EBSD investigation.     
}
\label{fig:ebsd2}
\end{figure}

To corroborate our conclusion of the absence of charge order with a $\mathbf{q}=(\pm 1/3,0)$ ordering vector in stoichiometric LaNiO$_{2}$ [Fig. 7 in the main text], we carried out additional RIXS measurements at different locations on the surface of the same crystal and on two additional crystals. A compilation of the data (quasielastic scattering integrated from $-50\leq E\leq 50$~meV) is presented in Supplementary Fig.~\ref{fig:Qmap_LV_3samples}. We find that irregular humps and dips occur in all data sets, apparently at  random $H$ values. The intensity of these humps and dips is comparable to that of the ``peak-like'' feature at $H\sim0.27$ in Fig. 7 of the main text [see also crystal 1, position 1 in Supplementary Fig.~\ref{fig:Qmap_LV_3samples}]. Presumably, these features of local humps and dips in the quasielastic scattering intensity originate from the uneven topography of the polished surfaces of the LaNiO$_{2}$ crystals, characterized by cracks and excavations [Supplementary Fig.~\ref{fig:Xtal} and Fig.~\ref{fig:optical}]. In contrast, well-defined peaks at symmetric $(\pm 1/3,0)$ positions \cite{RossiNatPhys2022} are not observed for our LaNiO$_{2}$ crystals. 

In addition, taking the intensity of the charge density wave (CDW) peak from a LaNiO$_{2}$ thin film sample reported  in Ref.~\cite{RossiNatPhys2022} as a reference, we estimate the expected CDW signal strength in our RIXS spectra of the LaNiO$_{2}$ crystal under the assumption that a comparable CDW would be present. To ensure direct comparability, we normalize RIXS spectra from our experiment and those in Ref.~\cite{RossiNatPhys2022} by the integrated intensity of the respective $dd$ excitations [see Supplementary Fig.~\ref{fig:CDW_estimation}(a)]. 
The result of this analysis is shown in Fig.~7(b) in the main text, suggesting that the presence of a CDW analogous to that reported in Ref.~\cite{RossiNatPhys2022} would lead to quasielastic scattering intensity around $H = 1/3$ in our spectra markedly higher than the irregular humps and dips visible in Supplementary Fig.~\ref{fig:Qmap_LV_3samples}. Note that for better compatibility we added an offset in vertical direction to the reference data from Ref.~\cite{RossiNatPhys2022} in Fig.~7(b) in our main text, while Supplementary Fig.~\ref{fig:CDW_estimation}(c) shows the data without offset. The higher background level in our LaNiO$_{2}$ crystal data is likely due to enhanced quasielastic scattering from the uneven topography of the crystal. 

Furthermore, we normalize RIXS spectra from a NdNiO$_{2}$ thin film study~\cite{TamNatMat2022} in Supplementary Fig.~\ref{fig:CDW_estimation}(b) and display the integrated quasi-elastic intensity in Supplementary Fig.~\ref{fig:CDW_estimation}(c). The samples NdNiO$_{2}$ film-1 and  NdNiO$_{2}$ film-2 were reduced using CaH$_2$ at 200 and 220$^\circ$C, respectively. Remarkably, the CDW signal from NdNiO$_{2}$ film-1 is highly intense compared to that of the other samples (note the scaling factor of 0.3).

\begin{acknowledgements}

\end{acknowledgements}


\bibliography{LaNiO2_RIXS}

\newpage

\begin{figure}[tb]
\includegraphics[width=1\columnwidth]{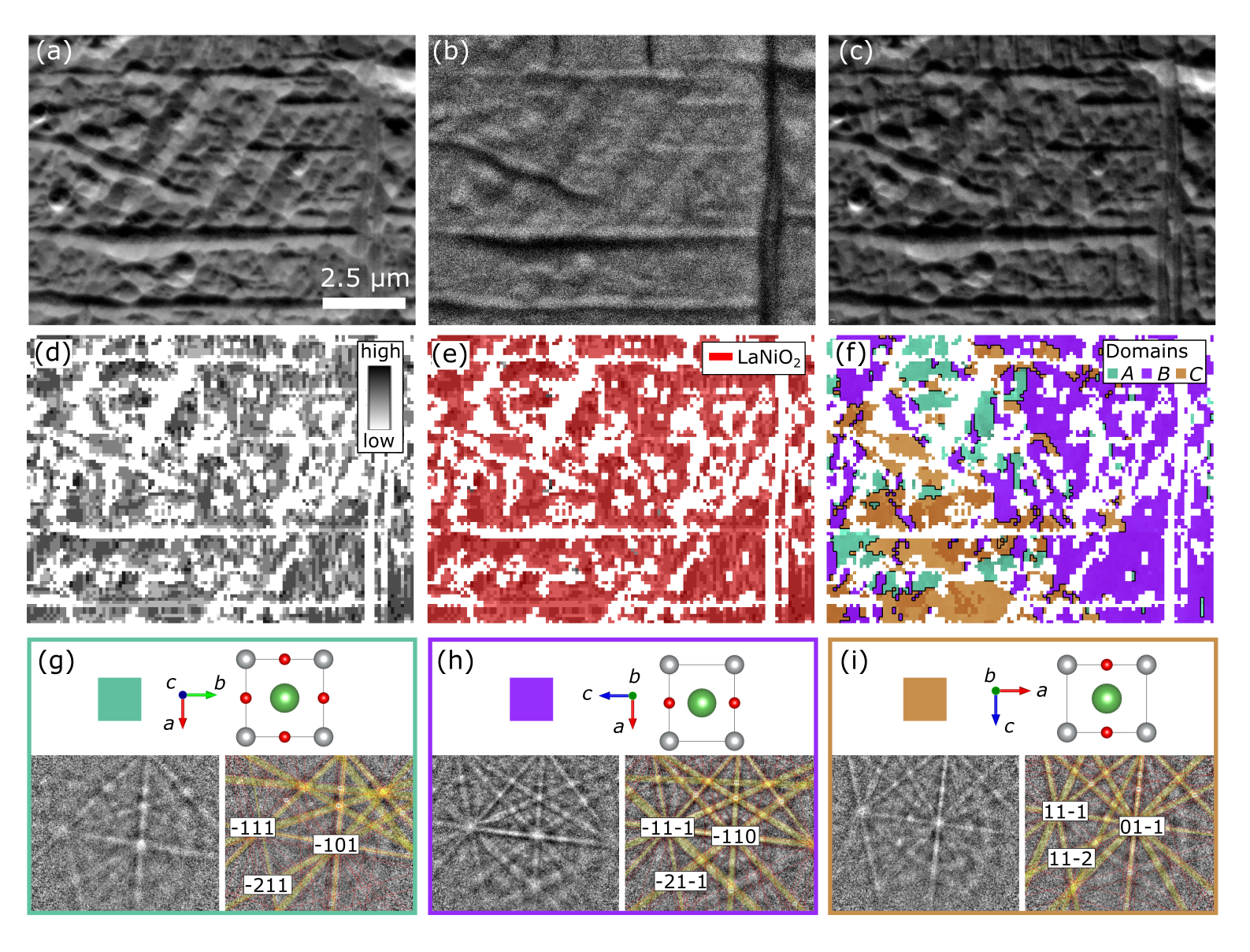}
\caption{SEM images and EBSD maps of the third surface region of the LaNiO$_{2}$ crystal  selected for the EBSD investigation.     
}
\label{fig:ebsd3}
\end{figure}


\begin{figure}[tbp]
\includegraphics[scale=1]{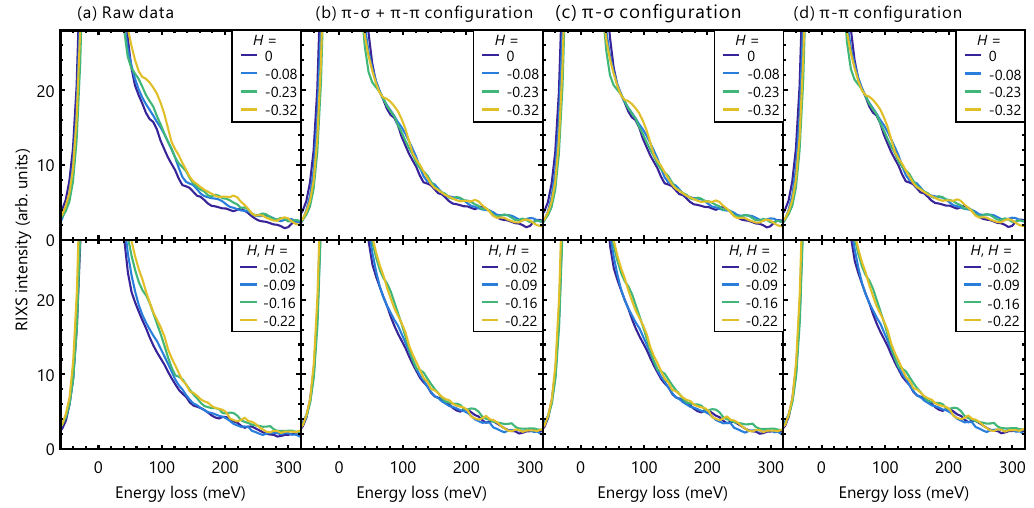}
\caption{Self-absorption correction of the selected RIXS spectra taken along the $(H,0)$ (top panel) and $(H,H)$ (bottom panel) directions. (a) The raw data before the correction. (b)-(d) Corrected intensities applying (b) an averaging over the $\pi$-$\sigma$ and $\pi$-$\pi$ configurations for incident and outgoing photons, (c) the $\pi$-$\sigma$ configuration, and (d) the $\pi$-$\pi$ configuration.}
\label{fig:self_absorption}
\end{figure}

\begin{figure}[tbp]
\includegraphics[scale=1]{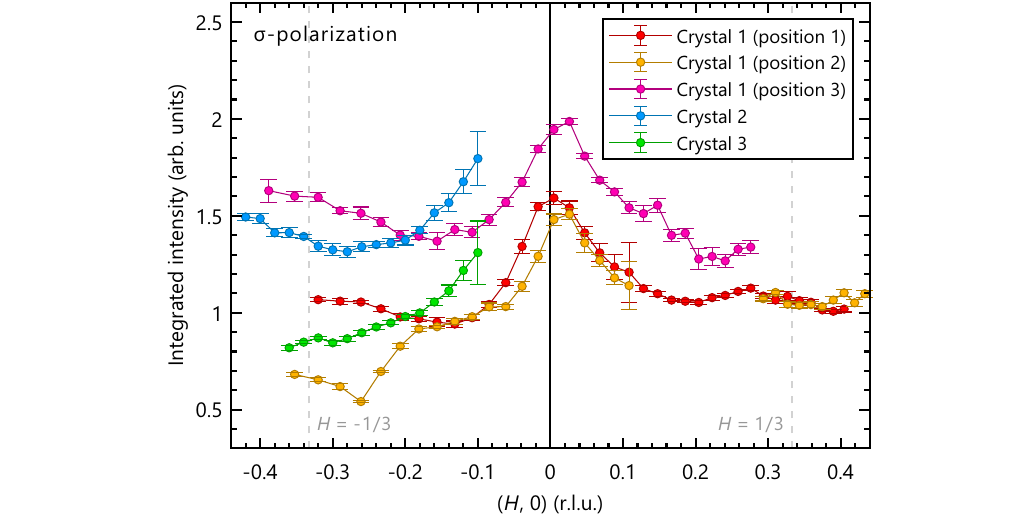}
\caption{Integrated intensity of quasielastic scattering of RIXS spectra along the $(\pm H,0)$ direction, acquired with $\sigma$ polarized photons from different LaNiO$_{2}$ crystals. The integration range is $-50\leq E\leq 50$~meV energy loss. Error bars are determined by the standard deviation of six RIXS spectra at each momentum. The red symbols are the data presented in Fig. 7 in  the main text (crystal 1), while the orange and purple symbols correspond to data obtained from distinct positions on the same crystal. The blue and green symbols represent data measured from different crystals.}
\label{fig:Qmap_LV_3samples}
\end{figure}

\begin{figure}[tbp]
\includegraphics[scale=1]{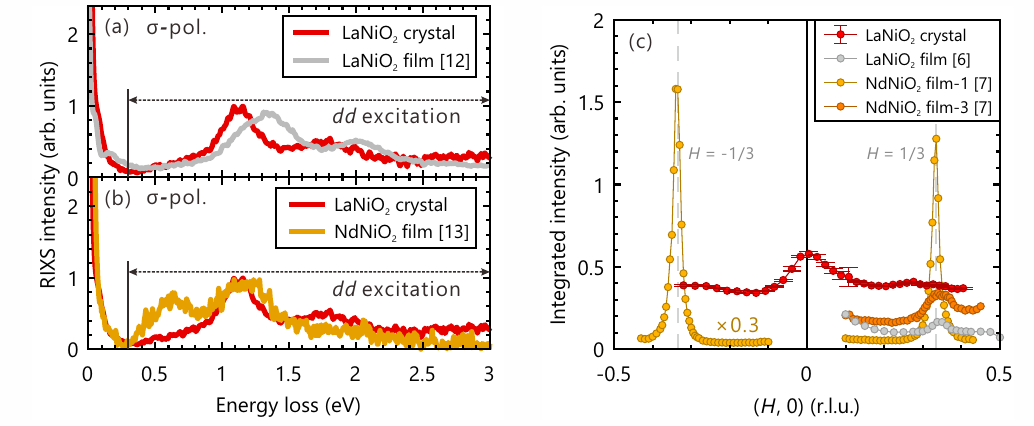}
\caption{Estimation of the expected CDW signal intensity for LaNiO$_{2}$ crystals based on RIXS reference data from thin films. (a) Normalization of the RIXS spectra of the LaNiO$_{2}$ crystal and a LaNiO$_{2}$ thin film~\cite{RossiNatPhys2022}, according to the integrated intensities of the respective $dd$ excitations in the range $0.3\leq E\leq 3$~eV. Both RIXS spectra were acquired with $\sigma$-polarized photons. The $\mathbf{q}$ positions are $(0.34,0)$ for the crystal, $(0.35,0)$ for the LaNiO$_{2}$ thin film~\cite{RossiNatPhys2022}. (b)  Normalization of the RIXS spectra of the LaNiO$_{2}$ crystal and a NdNiO$_{2}$ thin film~\cite{TamNatMat2022}. The $\mathbf{q}$ position of the NdNiO$_{2}$ thin film is $(-0.35,0)$. (c) Quasielastic scattering intensities along the $(\pm H,0)$ direction normalized by the integrated intensity of the $dd$ excitations for each sample. 
In addition, the data from NdNiO$_{2}$ film-1~\cite{TamNatMat2022} is multiplied by a factor of 0.3 to match the scale of the other samples.}
\label{fig:CDW_estimation}
\end{figure}
